\documentclass[superscriptaddress,showkeys]{revtex4} 
\usepackage{graphicx,amsmath,latexsym,amssymb}
\usepackage{color}
  
\usepackage[hidelinks]{hyperref}
\hypersetup{
  colorlinks   = true, 	
  urlcolor     = blue, 	
  linkcolor    = blue, 	
  citecolor   = magenta 	
}

\usepackage{lineno}
\usepackage{bm}
\newcommand{\mathsym}[1]{{}}

\usepackage{graphicx}
\newcommand{\be}{\begin{equation}}
\newcommand{\ee}{\end{equation}}
\newcommand{\bea}{\begin{eqnarray}}
\newcommand{\eea}{\end{eqnarray}}
\newcommand{\bd}{\begin{displaymath}}
\newcommand{\ed}{\end{displaymath}}

\begin{document}

\title{The spin-one Duffin-Kemmer-Petiau equation revisited: analytical study of its structure and a careful choice of interaction}

\author{M. Baradaran}
\email{marzieh.baradaran@uhk.cz}
\affiliation{Department of Physics, Faculty of Science, University of Hradec Kr\'alov\'e,
 Rokitansk\'eho 62, 500 03 Hradec Kr\'alov\'e, Czechia}
\author{L.M. Nieto}
\email{luismiguel.nieto.calzada@uva.es}
\affiliation{Departamento de F\'{\i}sica Te\'{o}rica, At\'{o}mica y \'{O}ptica, and Laboratory for Disruptive Interdisciplinary Science (LaDIS), Universidad de Valladolid, 47011 Valladolid, Spain}
\author{L.P. de Oliveira}
\email{oliveira.phys@gmail.com}
\affiliation{Instituto de Pesquisas Energéticas e Nucleares, IPEN/CNEN, Av. Prof. Lineu Prestes, 2242, Cidade Universitária, CEP 05508-000 São Paulo, SP, Brazil}
\author{S. Zarrinkamar}
\email{saber.zarrinkamar@uva.es}
\affiliation{Departamento de F\'{\i}sica Te\'{o}rica, At\'{o}mica y \'{O}ptica, and Laboratory for Disruptive Interdisciplinary Science (LaDIS), Universidad de Valladolid, 47011 Valladolid, Spain}
\affiliation{Departament of Physics, Garmsar Branch, Islamic Azad University, Garmsar, Iran}

\date{\today}

\begin{abstract}
The Duffin-Kemmer-Petiau equation is investigated for spin one bosons with the so-called natural (normal) and unnatural (abnormal) parity states for non-minimal vector interactions. 
To illustrate the current state of knowledge about the equation, a  thorough but concise discussion is made on what can be achieved analytically  within this framework for well-known phenomenological interactions, including Coulomb, soft-core, Cornell, Kratzer, and exponential type interactions. In the non-exponential cases, the equation, depending on the chosen interaction, is studied in relation to the confluent, doubly-confluent, and biconfluent Heun functions. 
Furthermore, to show the need for careful treatment of various parity states, a Kratzer-type potential, such as a generalized Coulomb interaction, is discussed in depth using the Lie algebraic approach, showing the need for careful analysis of abnormal parity states in a fairly explicit way. 
The energies obtained are discussed using some figures to explicitly show the different regimes, as well as the absence of the Klein paradox. 
Finally, some directions for future work that would undoubtedly need to be explored in this field are discussed.
\end{abstract}

\date{\today}


\maketitle

\section{Introduction}

In relativistic quantum mechanics, the Klein-Gordon, Dirac, and Proca-Weinberg equations describe, respectively, spin-zero, spin-12, and spin-one particles, and the Duffin-Kemmer-Petiau (DKP) equation provides a framework within which both spin-zero and spin-one particles can be investigated \cite {Petiau,Kemmer,Duffin,Greiner}.
The case of spin zero, due to a  structure similar to the Klein-Gordon equation, has been frequently studied in the literature \cite {Greiner, Dong}. 
The spin-one version of the equation becomes more interesting when we remember the complicated nature of the Proca and Weinberg equations, as well as the scant literature that exists on them.
On the other hand, the DKP equation, due to the way it is represented, which appears in the form of a ten-spinor wave function, is a rich basis for analyzing various interactions and fields.
The structure of this equation has motivated recent papers investigating the problem with a variety of physical interactions and concepts: for example,
Nedjadi and Barret \cite {Nedjadi 1994} considered the free particle, potential well, and Coulomb problem for vector bosons, solutions for DKP with a linear interaction were found in \cite {Cardoso 2010}, and the DKP oscillator was analyzed within a noncommutative approach in \cite {Hassanabadi 2012}. 
 The similarity of the structures in DKP and Dirac equations was discussed in the conceptual work of Okni\'nski \cite{Okninski} and the creation rate within the de Sitter framework was obtained using the Bogoliubov transformation in \cite{Kangal}. 
Castro and de Oliveira provided a review of the equation with non-minimal vector interactions, including the linear one \cite {Castro 2014},
and the Galilean DKP theory was formulated at finite temperature to study Bose-Einstein condensation \cite {Bose}.
An instructive survey on the equivalence of the DKP equation with the Klein-Gordon and Proca equations is provided in \cite {Castro PRA}. 
In \cite {Zarrinkamar 2016} ground state solutions of a class of generalized Coulomb interactions were derived using the Ansatz approach.
The fractional Hamiltonian formulation of the equation in the Lagrangian formalism was restudied by Bouzid and Merad \cite {Bouzid 2017}. 
The equation in (1+1)-dimensions was analyzed in \cite {Lunardi 2017}.
The associated Aharanov-Bohm problem was discussed in detail by Castro and Silva \cite{Castro 2018}, and the field-theoretical analysis of the equation was formulated in \cite{Castro 2018 2}, a work in which the corresponding Stefan-Boltzmann law was also discussed, as well as the Casimir effect at finite temperature.
Later, de Montigny and Santos \cite{deMontigny} generalized the formulation of the equation to arbitrary dimensions.

Solutions of the DKP equation with a Kratzer potential in a noncommutative approach were obtained in \cite{Saidi}, a work in which some delicate points should have been carefully considered, as we will see in some of the following sections.
Let us also mention that the non-relativistic limit of the equation was found in a non-commutative framework using the Foldy-Wouthuysen transformation and the Moyal-Weyl product \cite{Haouam}.
The Dirac delta problem in the $q$-deformed formulation of the equation was solved by Sobhani et al. \cite {Sobhani}, and using Dirac derivatives, Chargui and Cherif found the DKP equation valid within a $\kappa$-Minkowski spacetime \cite{Chargui 2020}.
Increasing the spatial dimensions, the $(2+1)$-dimensional DKP oscillator influenced by an external magnetic field and the corresponding canonical ensemble thermodynamic properties were reviewed in \cite {Gomez}.
On the other hand, the study of the momentum space of the three-dimensional DKP oscillator within the Snyder-de Sitter framework was performed in \cite{Hamil 2021}.
In a conceptual paper, Chragui and Dhahbi analyzed an extended DKP oscillator for spin-one bosons, exhaustively studying various parity states \cite{Scripta}, and subsequently treated the DKP oscillator in three spatial dimensions in a noncommutative plane, showing the salient related physical aspects \cite{Chargui 2023}.

In recent years there has been intense activity in this area, with the following results being noteworthy: the rotational symmetries of the equation within the $(1+2)$-dimensional G\"urses spacetime background were commented on by Candemir and Ahmed \cite {Candemir 2023}, the equation with a hyperbolic tangent potential was investigated by Valladares and Rojas \cite {Valladares}, the DKP oscillator problem in one and two spatial dimensions and the associated thermal properties were analyzed within a noncommutative framework \cite{Moussa 2023}, a new non-minimal coupling associated with a $q$-deformed approach was considered within the framework of the equation and the thermodynamic properties of the $q$-deformed oscillator were studied \cite{Boumali 2023}, and Chargui et al. generalized the DKP oscillator in a form that includes an additional spin-orbit term \cite{Chargui res}, to mention only the most prominent ones. 
To conclude this extensive description of results we will mention that although the DKP equation has been frequently used in particle and nuclear physics, there are recent applications of the equation in other areas of physics \cite {Menezes}.
The motivation for such studies is not exceptional because spin-one systems are present in many areas, including molecular systems \cite{Parks, Albayrak}, isotropic spin-1 antiferromagnetic systems \cite{Nag}, Cooper pairs \cite{Cheng}, charged impurities \cite{Vazquez}, etc.
Very recently, Charqui and Dhahbi \cite{JC} have addressed a very interesting and applied aspect of the DKP equation: they have connected the equation in two spatial dimensions in a kind of atom-field interaction similar to the Jaynes-Cummings model, which may be a great step in the field if we remember the limitations of other models. However, this idea is quite recent and absolutely open to debate, with many questions and challenges that need to be properly answered.

The present work focuses on two main points: (i) a general analysis of the structure of the spin one DKP equation and what can be derived analytically  from a fairly general class of phenomenological interactions and (ii)
a clear and detailed discussion of so-called unnatural or abnormal parity states, which, as far as we know, has not been done before. 
To this end, we start in Section~\ref{DKPstructure} from the ordinary differential equations governing the DKP spinor components to avoid extensive text citations of previous work. 
The reader must either be familiar with the spin one DKP equation, or carefully read the original works on the derivation of various parity states \cite {Nedjadi 1994, Cardoso 2010, Castro 2014, Scripta} in order to understand the  concepts  and calculations. 
We next present the Lie-algebraic approach to tackle the equations arising in the case of a carefully chosen Kratzer-type interaction. 
We derive the general solution in Section~\ref{QESgen} and analyze the energies for both the natural and unnatural parts to provide a clear discussion on the physics of the equation. 
In Section~\ref{Heun}, without going into details, the forms of the equation with a large class of phenomenological interactions are discussed in terms of Heun functions and other approaches in mathematical physics. 
Concluding remarks are included in the last section.

\section{The Structure of DKP Equation}\label{DKPstructure}

Before starting with the essential mathematical developments, let us mention some points about the limitation in the analytical study of the spin one DKP equation.
To avoid overlaps with previous publications, we will make some statements and refer to the original articles \cite {Nedjadi 1994, Cardoso 2010, Castro 2014} to support them.
A look at equations (61) and (62) of \cite {Nedjadi 1994} reveals that the general form of the equation appears as a system of coupled second-order differential equations that, to our knowledge, has not been solved analytically and surely it cannot be solved. 
One might think that the homogeneous part of Eq. (61) in \cite {Nedjadi 1994} is the general form of the Heun equation \cite {Ronveaux} and that, in this case, the solution can be reported. 
However, this is not true since the spinor components that appear there are coupled and we cannot simply think of a single dependent variable problem. 
This very complicated form of the equation has forced the authors to focus on the case of $J=0$ in order to report the analytical solutions or, as will be seen soon, to consider a zero term in part of the interaction \cite {Nedjadi 1994, Cardoso 2010, Castro 2014}.

The components of the spin one DKP equation are denoted as the parity states $(-1)^j$ and $(-1)^{j+1}$.
This terminology originates from the behavior of the associated spherical harmonics of each spinor under Maxwell-like equations.
Those that retain their value of $J$ are called $(-1)^j$ components, or natural/normal/magnetic-like parity states, and those with a one-unit increase or decrease in the value of their $J$ are called $(-1)^{j+1}$ components, or unnatural/abnormal-electric-like parity states. 
The connection to electric and magnetic fields comes simply from the behavior of these fields under parity, as they have negative and positive parities, respectively. 
Let us now review the form that the final equations take and emphasize that the reader who is not an expert on the subject should first read the articles \cite {Nedjadi 1994, Cardoso 2010, Castro 2014, Scripta}.

\subsection{$(-1)^j$--Parity Components}

As already mentioned above, we work on a ten-component basis, in which, roughly speaking, we denote by $F_0(r)$ the term related to the original component of parity preserved through some relations. Once $F_0(r)$ is determined, all data related to natural parity states are derived. After a little algebra it follows that this function must satisfy the differential equation \cite {Nedjadi 1994, Cardoso 2010, Castro 2014}
\begin{equation}\label{F}
\frac{d^2F_0(r)}{dr^2}+\left[E^2-M^2-\frac{dA_r}{dr}-\frac{j(j+1)}{r^2}+A_0^2-A_r^2\right]F_0(r)=0,
\end{equation}
where $j$ denotes the order of the associated spherical harmonics.
The term $A_r(r)$ is a spherically symmetric potential and the scalar part of the original DKP equation. 
We consider $A_r(r)$ and $A_0(r)$ in the form of the modified Kratzer potential $V_{MK}=D_e(\frac{r-r_e}{r})^2$, with $D_e>0$ \cite{Kratzer}, indeed
\begin{equation}\label{Kratzer}
A(r)=\frac{x_r}{r}+\frac{y_r}{r^2}+z_r, \ \ \ \ \
A_0(r)=\frac{x_0}{r}+\frac{y_0}{r^2}+z_0,
\end{equation}
In this form, $x_r<0,\ y_r>0$ and $z_r>0$. 
Substitution of the potential \eqref{Kratzer} in \eqref{F} yields
\begin{equation}
	\begin{aligned}
\bigg\{\frac{d^2}{dr^2}+E^2-M^2+z_0^2-z_r^2&+\frac{2(x_0z_0-x_r z_r)}{r}+\frac{2(y_0 z_0-y_r z_r)-j(j+1)+x_r+x_0^2-x_r^2}{r^2}\\
&\, +\frac{2(x_0 y_0-x_r y_r + y_r)}{r^3}+\frac{y_0^2-y_r^2}{r^4} \bigg\}F_0(r)=0.
	\end{aligned}
	\label{edof0}
\end{equation}
Let us now review the equations governing the unnatural parity states.

\subsection{$(-1)^{j+1}$--Parity Components}
In this case, in order to report analytical solutions, we have to consider $A_0=0$. 
In this case, we have to consider two equations for the spinor components \cite {Nedjadi 1994, Cardoso 2010, Castro 2014}:
\begin{equation}\label{H}
\frac{d^2H_0}{dr^2}+\left[E^2-M^2+\frac{dA_r}{dr}-\frac{j(j+1)}{r^2}-A_r^2\right]H_0=0
\end{equation}
and
\begin{equation}
\label {phi}
\frac{d^2 \phi}{dr^2}+\left[E^2-M^2-\frac{dA_r}{dr}-\frac{j(j+1)}{r^2}-\frac{A_r}{r}-A_r^2\right]\phi=0,
\end{equation}
which take the following forms for the potential \eqref{Kratzer}:
\begin{equation}\label{edoh0}
\left\{\frac{d^2}{dr^2}+E^2-M^2-z_r^2-\frac{2x_r z_r}{r}-\frac{j(j+1)+x_r(x_r+1)+2y_r z_r}{r^2}-\frac{2(x_r +1) y_r}{r^3}-\frac{y_r^2}{r^4}\right\}H_0(r)=0,
\end{equation}
and
\begin{equation}\label{edophi}
\left\{\frac{d^2}{dr^2}+E^2-M^2-z_r^2-\frac{(2x_r+1) z_r}{r}-\frac{j(j+1)+x_r^2+2y_r z_r}{r^2}+\frac{(1-2x_r ) y_r}{r^3}-\frac{y_r^2}{r^4}\right\}\phi(r)=0.
\end{equation}
In the next section we will look for analytical solutions to the three equations \eqref{edof0}, \eqref{edoh0} and \eqref{edophi}.

\section{Quasi-Polynomial Solutions}
\label{QESgen}
First we review the basic concepts of quasi-exact solubility \cite{Turbiner 1988} and then we will comment on the solutions and the spectrum associated with each of the three differential equations obtained above.

 \subsection{Algebraic Structure with Kratzer-like Interaction}
It is clear that the three equations \eqref{edof0}, \eqref{edoh0} and \eqref{edophi} appear in the form of Schr\"odinger equations $H\psi(r)=0$ with Hamiltonians of the same type 
\begin{equation}\label{form}
H=\frac{d^2}{dr^2}+\alpha+\frac{\beta}{r}+\frac{\omega}{r^2}+\frac{\delta}{r^3}+\frac{\sigma}{r^4},
\end{equation}
already analyzed in \cite{Turbiner 1988, Turbiner 2016, Gonzalez}.
Applying the transformation
\begin{equation}\label{waveQES}
\psi(r)=r^{1-\eta} \exp\left[-\sqrt{-\alpha}r-\frac{\sqrt{-\sigma}}{r}\right]R(r), \qquad \eta=\frac{\delta}{2\sqrt{-\sigma}},
\end{equation}
in which $1-\eta>0$ and $\sigma,\alpha<0$, Eq. (8) can be written in the form $\tilde H R(r)=0$ with 
\begin{equation}
\tilde H=r^2\frac{d^2}{dr^2}-2\left(\sqrt{- \alpha}r^2-(1-\eta)r-\sqrt{-\sigma}\,\right)\frac{d}{dr}-(ar+b),
\end{equation}
where
\begin{equation}
a=2(1-\eta)\sqrt{-\alpha}-\beta, \qquad\qquad \ b=\eta(1-\eta)+2\sqrt{\alpha \sigma}-\omega.
\end{equation}
The differential operator $\tilde H$ can be written as an element of the universal enveloping algebra of $sl(2)$ as \cite{Turbiner 1988, Turbiner 2016, Gonzalez}, 
\begin{equation}\label{Lieform}
\tilde H_{QES}=-\mathcal{J}_n^+ \mathcal{J}_n^- +2\sqrt{-\alpha}\mathcal{J}_n^+ -2\sqrt{-\sigma}\mathcal{J}_n^- +(2\eta-n-2)\left(\mathcal{J}_n^0+\frac{n}{2}\right)+b,
\end{equation}
provided that
\begin{equation}\label{qesCons}
a=-2n\sqrt{-\alpha},  \quad n=0,1,2,\dots
\end{equation}
In \eqref {Lieform} the generators of the algebra $sl(2)$ are \cite {Turbiner 1988}
\begin{equation}\label{GenerJJJ}
		\mathcal{J}_n^+ =r^2\,\frac{d}{dr}-n\,r ,\qquad
		\mathcal{J}_n^0 = r\,\frac{d}{dr}-\frac n2,\qquad
		\mathcal{J}_n^- = \frac{d}{dr},  \quad n=0,1,2,\dots,
\end{equation}
and satisfy the commutation relations \cite {Turbiner 1988}
\begin{equation}
[\mathcal{J}^+,\mathcal{J}^-]=-2\mathcal{J}^0, \ \ \ \ \ \  [\mathcal{J}^{\pm},\mathcal{J}^0]={\mp}\mathcal{J}^{\pm},
\end{equation}
that leave invariant the $(n+1)-$dimensional linear space of finite order polynomials $P_{n+1}(r)=\langle1,r,r^2,...,r^n\rangle$. As a result, $H_{QES}$ preserves the finite-dimensional space of polynomials
\begin{equation}
R(r)=\sum_{m=0} ^n{c_m r^m},
\end{equation}
and we may finally write he wave-function \eqref{waveQES} as \cite{Turbiner 1988}
\begin{equation}\label{wave}
\psi_n(r)=r^{1-\eta} \exp\left[-\sqrt{-\alpha}r-\frac{\sqrt{-\sigma}}{r}\right] \sum_{m=0} ^n{c_m r^m},
\end{equation}
where the coefficients $c_m$ satisfy the three-term recursion relation
\begin{equation}\label{recursion}
c_{m+1}=\frac{\left(b+m(2\eta-m-1)\right)c_m +2(m-n-1) \sqrt{-\alpha}\, c_{m-1}}{2(m+1)\sqrt{-\sigma}},
\qquad \text{with}\qquad c_{-1}=c_{n+1}=0.
\end{equation}
This yields a system of linear equations which has nontrivial solutions if \cite {Turbiner 1988, Panahi 2016 2, Baradaran 2018}  
 \begin{equation}\label{matrix}
\left|
\begin{array}{ccccc}
	b & -2\sqrt{-\sigma} &   &   &   \\ [1ex]
	-2n\sqrt{-\alpha} &b-2(1-\eta) & -4\sqrt{-\sigma} &   &   \\  [1ex]
	  & -2(n-1)\sqrt{-\alpha} &\ddots & \ddots &   \\ [1ex]
	  &   & \ddots & \ddots & -2 n\sqrt{-\sigma} \\ [1.5ex]
	  &   &   & -2\sqrt{-\alpha} & b-n(n-2\eta+1)
\end{array}
\right| =0.
\end{equation}
This last condition imposes severe restrictions between the parameters that appear in the potential. It should be noted that the equations \eqref{edof0}, \eqref{edoh0} and \eqref{edophi} appear as  doubly-confluent Heun equations \cite{Ronveaux, El-Jaick}.
Now we are going to consider in detail the solutions to each of these three differential equations, separating them according to the parity of the component being considered.

\subsection {$(-1)^{j}$--Parity Solution}
Let us consider first the natural parity states. Using the previous section, and comparing  \eqref{edof0} and \eqref{form}, we deduce 
\begin{equation}
	\begin{aligned}
		\alpha&=E_n^2-M^2+z_0^2-z_r^2,\qquad\quad
		\beta=2(x_0z_0-x_rz_r),\\
		\omega&=2(y_0z_0-y_rz_r)-j(j+1)+x_r+x_0^2-x_r^2,\\
		\delta&=2(x_0 y_0 - x_r y_r+y_r),\qquad\quad
		\sigma=y_0^2-y_r^2,\\
	\end{aligned}
\end{equation}
from which, together with \eqref{qesCons}, the general expression of the eigenenergies is deduced, which can be rewritten as
\be\label{energynormal1}
E_n=\pm\sqrt{M^2-\epsilon_n^2}, \quad \text{with}\quad
\epsilon_n^2\equiv (y_r^2-y_0^2)\left(\frac{x_rz_r-x_0z_0}{x_0y_0-x_ry_r+y_r-(n+1)\sqrt{y_r^2-y_0^2}}\right)^2-z_r^2+z_0^2,\quad  n=0,1,2,\dots
\ee
\begin{figure}[htb]
\centerline{\includegraphics[scale=.31]{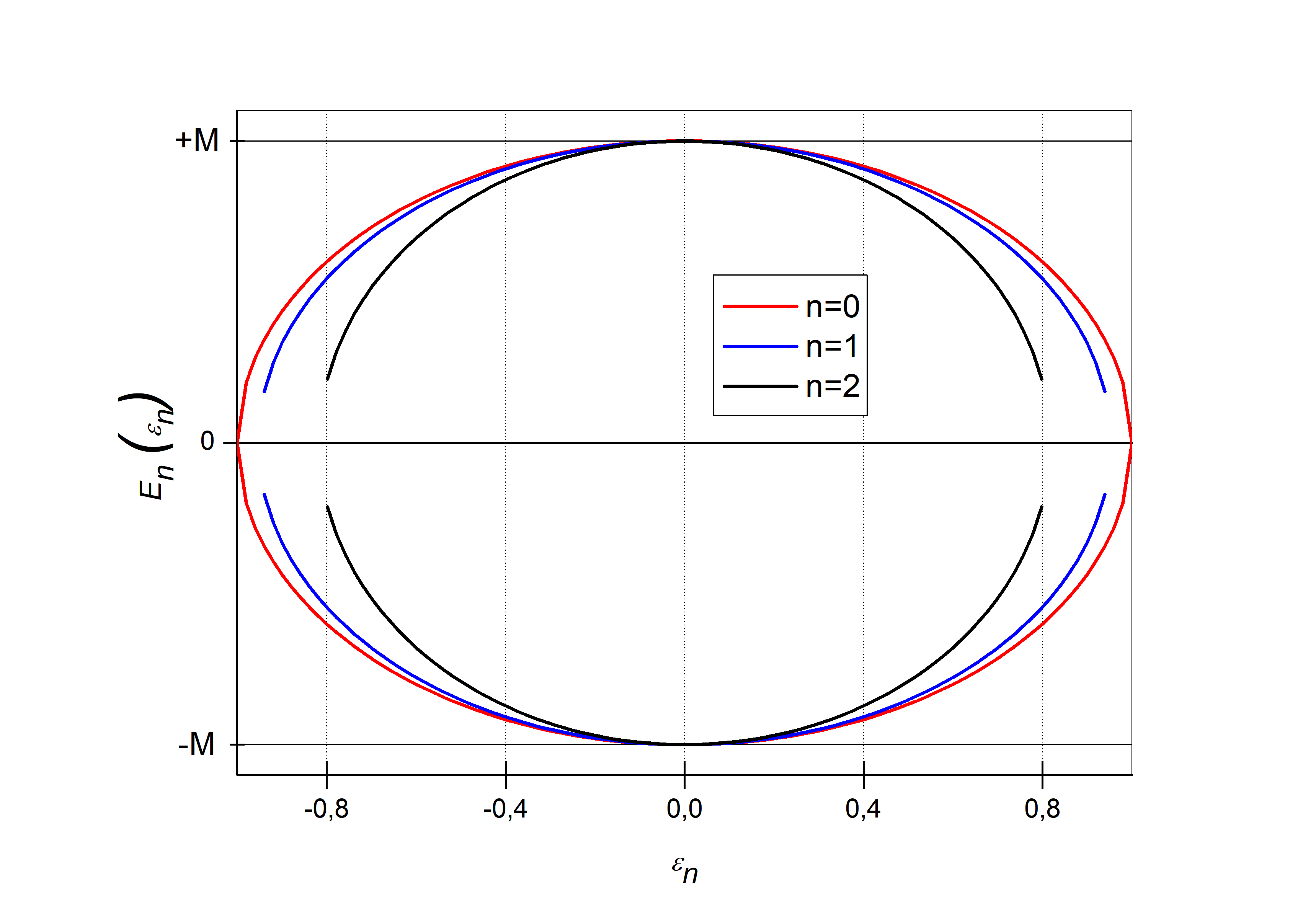}}
\caption{Energy behavior for normal parity states.}
\label{normal}
\end{figure}
In Figure~\ref{normal} you can see a graph of \eqref{energynormal1} for the three lowest values of $n$.
The energy levels are represented in the real plane $(E_n,\epsilon_n)$ for $M^2>{\epsilon_n^2}$, which implies $|y_r|>|y_0|$.
The families of curves are symmetric at $E_n=0$, implying that even for finite potentials, energy levels arise in the particle continuum and penetrate into the antiparticle region. 
In this way, there is no room for Klein's paradox. We can also observe that for $|y_0|>|y_r|$, the eigenenergies gain a phase in the radial component.

For non-minimal vector interactions, Cardoso et al. had already demonstrated properties of the DKP equation that inhibit the occurrence of the Klein paradox in $(1+1)$-dimensions, that is, the spatial component of the potential plays a determining role in the confinement of bosons \cite{Cardoso Klein}. 
Therefore, our work shows that the same can happen in non-minimal vector potentials in $(3+1)$-dimensions.

Regarding the relevant component in this case, from \eqref{wave} we obtain that its expression is
\begin{equation}\label{wavenormal}
F_{0,n}(r)=r^{1-(x_0y_0-x_ry_r+y_r)/\sqrt{y_r^2-y_0^2}}\ \exp\left[-r\sqrt{M^2-E_n^2+z_r^2-z_0^2}-\frac{\sqrt{y_r^2-y_0^2}}{r}\right] \sum_{m=0}^n {c_mr^m}.
\end{equation}
Note that as we mentioned above, we must have $1-\eta>0$, which specifically means that
\begin{equation}
 \eta=\frac{x_0y_0-x_ry_r+y_r}{\sqrt{y_r^2-y_0^2}}<1.
\end{equation}
Moreover, note that the potential parameters in \eqref{energynormal1} and \eqref{wavenormal} must fulfill the constraint given by \eqref{matrix}. Let us now consider the unnatural parity solutions.

\subsection {$(-1)^{j+1}$--Parity States}
As we have already seen, we have to consider two different types of spinor components: those of $H_0(r)$ coming from \eqref{edoh0} and those of $\phi(r)$, coming from \eqref{edophi}. 
We first consider the equation governing $H_0(r)$.

\subsubsection  {$H_0(r)$ Solutions}

\begin{figure}[htb]
\centerline{\includegraphics[scale=.31]{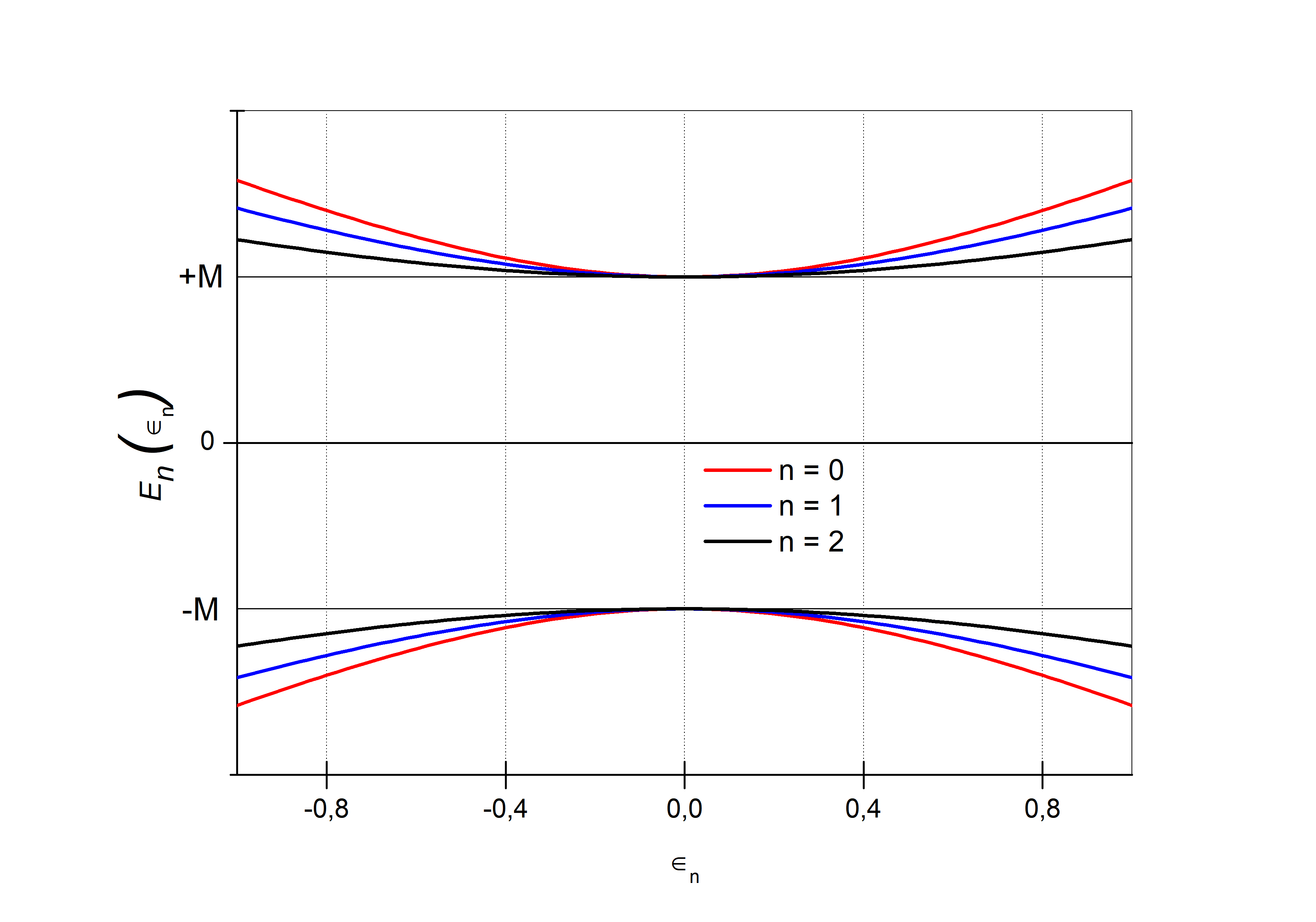}}
\caption{Energy behavior for abnormal parity states.
}
\label{abnormal}
\end{figure}

To obtain the solutions for $H_0(r)$ we have to compare \eqref{edoh0} and \eqref{form}, from which we deduce 
\begin{equation}
	\begin{aligned}
		& \alpha=E_n^2-M^2-z_r^2, \quad 
		\beta=-2x_rz_r,\quad \\
		&\omega=-\left(j(j+1)+x_r(x_r+1)+2y_rz_r\right)\quad\\
	&	\delta =-2 (x_r+1)y_r,\quad 
		\sigma=-y_r^2,
		\end{aligned}
\end{equation}
from which, together with \eqref{qesCons}, the possible values of the energy are deduced:
\begin{equation}\label{Henergy}
E_n^2=-\left(\frac{x_rz_r}{x_r+n+2}\right)^2+M^2+z_r^2, \qquad
 n=0,1,2,\dots
\end{equation}
These eigenenergies can written as 
\be\label{energyabnormal1}
E_n=\pm\sqrt{M^2+\epsilon_n^2}, \qquad \text{where}\qquad 
\epsilon_n^2=z_r^2 - \left( \frac{x_rz_r}{x_r+n+2}\right)^2, \qquad
n=0,1,2,\dots
\ee
In this case, energy levels do not penetrate the $E<|M|$ region, so there are no bound states and Klein's paradox. 
In Figure~\ref{abnormal} you can see a graph of \eqref{energyabnormal1} for the three lowest values of $n$.
Furthermore, the corresponding expression for the $H_{0,n}(r)$ components is obtained from \eqref{wave}:
\begin{equation}\label{Hwave}
H_{0,n}(r)=r^{x_r+2} \exp\left[-r\sqrt{M^2-E_n^2+z_r^2}-\frac{y_r}{r}\right] \sum_{m=0}^{n}c_mr^m.
\end{equation}
Recall that the potential parameters in both relations \eqref{energyabnormal1} and \eqref{Hwave} must satisfy the constraint given in \eqref{matrix}.

\subsubsection  {$\phi(r)$ Solutions}

In a completely analogous way we determine the values of the parameters in the case of the $\phi(r)$ component, comparing \eqref{edophi} and \eqref{form}, 
\begin{equation}
	\begin{aligned}
		\alpha&=E_n^2-M^2-z_r^2,\qquad
		\beta=-(2x_r+1)z_r,\\
		\omega&=-\left(j(j+1)+x_r^2+2y_rz_r\right)\\
		\delta&=(1-2x_r)y_r,\qquad
		\sigma=-y_r^2,\\
	\end{aligned}
\end{equation}
which, together with \eqref{qesCons}, give the energy relation as
\begin{equation}\label{phienergy} 
(2x_r+1)z_r=-(2n+2x_r+1)\sqrt{M^2-E_n^2+z_r^2},
\end{equation}
which does not have a real energy spectra, being the spinor component 
\begin{equation}\label{phiwave}
\phi_n(r)=r^{x_r+\frac{1}{2}} \exp\left[-r\sqrt{M^2-E_n^2+z_r^2}-\frac{y_r}{r}\right] \sum_{m=0}^{n}c_mr^m.
\end{equation}

Note that, as in the previous cases, the potential parameters in Eqs. \eqref{phienergy} and \eqref{phiwave} must satisfy the constraint given in \eqref{matrix}.

\section{Heun Forms and a class of Interactions}
\label{Heun}

The Heun differential equation \cite {Ronveaux} can be considered a generalization of other special functions in mathematical physics. This equation can be written in a form that resembles Schr\"odinger's, namely
$ \psi''(r)+V(r)\psi(r)=0$, with
\begin{equation}
V(r)=\frac{A_1}{r}+\frac{A_2}{r-1}+\frac{A_3}{r-a}+\frac{A_4}{r^2}+\frac{A_5}{(r-1)^2}+\frac{A_6}{(r-a)^2},
\end{equation}
in which all parameters except $r$ are constants chosen simply to show the form of the equation.
As already said, special cases of this equation produce other special functions of mathematical physics, including hypergeometric ones, this being a topic of current interest \cite {Ronveaux, El-Jaick, Ishkhanyan 2016}.
In the next subsection, we give the general idea by which some important interactions of physics can be  investigated analytically in terms of Heun functions or other simpler special functions. 
However, it is worth mentioning that each interaction can be the subject of an extensive study separately, since both the mathematical  and physical structure of the problem vary from one field to another.

\subsection{Generalized Cornell Interaction and Bi-confluent Heun Equation}

The Cornell potential includes linear and Coulomb terms. This is an attractive case in particle physics since it has a confining linear term in addition to the Coulomb interaction, and has been used to study bound states \cite{Eichten}. 
In its basic form, the resulting equation in the DKP formalism can be  investigated simply as it appears in the form of a non-relativistic three-dimensional harmonic term, the solutions of which have been known for decades \cite{Harmonic PRA, Dong 2007, Junker}. 
The generalization of the interaction with a  constant extra term appears in the form of bi-confluent Heun function with the effective Hamiltonian \cite{Ronveaux, El-Jaick} 
\begin{equation}
V(r)=A_1r+A_2r^2+\frac{A_{-1}}{r}+\frac{A_{-2}}{r^2}+k.
\end{equation}
It is interesting to note that the linear term is normally called the DKP oscillator, as it results in a harmonic-like form due to the square term, resembling the non-relativistic harmonic oscillator.

\subsection{Coulomb plus Soft-Core Coulomb Interaction and Confluent-Heun Equation}

Mehta, a long time ago, proposed a change in the denominator of the Coulomb interaction \cite {Mehta}, which is called the truncated or soft-core Coulomb problem.
Although the soft-core potential was originally used in particle and nuclear physics, its applications now go much further and the potential is of great interest and application in laser physics, optics, etc. \cite{Revmodphys, PRR Young}. 
The combination of ordinary and soft-core  Coulomb terms, that is, the interaction $\frac{a_{-1}}{r}+\frac{A_{-1}}{r-\beta}$, takes the form of confluent-Heun differential equation   \cite{Ronveaux, El-Jaick, Ishkhanyan 2016}, with
\be
V(r)=\frac{A}{r}+\frac{B}{r-1}+\frac{C}{r^2}+\frac{D}{(r-1)^2}+k.
\ee
The soft-core Coulomb interaction can also be investigated on its own within this approach.

\subsection{Kratzer-like Interaction and Double-confluent Heun Equation}
In the previous section of this work we have investigated the Kratzer-type potential with the Lie algebraic approach.
However, the problem can also be  investigated in terms of the double-confluent Heun functions \cite {Ronveaux, El-Jaick, Ishkhanyan 2016}. 
By making a simple change of variable, the radial equations with our Kratzer-like interaction, appear in the equivalent form of the doubly-confluent Heun equation, with 
\begin{equation}
	V(r)=\frac{A_{-1}}{r}+\frac{A_{-2}}{r^2}+\frac{A_{-3}}{r^3}+\frac{A_{-4}}{r^4}+k.
\end{equation}
whose solutions can be written in the form \cite{Ronveaux, El-Jaick, Ishkhanyan 2016}

\subsection{Yukawa and Exponential-type Interactions}
In dealing with exponential-type interactions, it is not possible to present analytical solutions unless we make some approximations. 
Obviously, this is due to the simultaneous presence of inverse and exponential terms.
One way to report general solutions is the so-called Pekeris-type approximation, which considers some exponential approximation for the inverse terms, or vice versa \cite {Pekeris, aldrich}. 
However, this is very rough since the approximation works  only for a very limited range and can not be very competitive with powerful numerical techniques. 
Nevertheless, there are papers that report analytical solutions with such approaches for various equations of quantum mechanics.

\section{ Conclusion}\label{Conclusion}

The spin one DKP equation, unlike other wave equations in quantum mechanics, has not been  discussed extensively in the literature. 
There may be several reasons for this, including the rather complicated structure of the equation, which usually appears in the form of ten components. 
The analysis of the DKP equation in this case must be done very  carefully due to the existence of the so-called unnatural or abnormal parity states. 
In the work we have presented, to the best of our knowledge, we considered for the first time  a Kratzer interaction in which an imaginary energy is observed in the case of a component with unnatural parity, explicitly demonstrating the need to  consider all parity states. 
Furthermore, the energy of the natural parity and unnatural parity cases shows the relationship between the natural and unnatural states with $S=0$ and $|S|=1$. 
To solve the DKP equation with this interaction we have used the powerful Lie algebraic approach and report the general solution.

We also discuss other classes of interactions, including the Cornell, soft-core, exponential and trigonometric,  which are not only pedagogically important to graduates, but also highly physically motivating and widely applicable in various fields.
We also comment on the energy of various parity states that may, in some way, indicate the absence of  Klein's historical paradox.

We hope that with the fairly detailed conceptualization and calculations performed here, the DKP equation will be  considered as a basis for studying spin one ions, molecules, hadrons, etc. 
As a final point, let us mention that it seems that the analysis of the non-Hermitian structure of the equation is very deficient in the field and studies in this direction could open quite new and deep horizons.

\section*{Acknowledgments}
The work of M.B. was supported by the Czech Science Foundation within the project 22-18739S.
The research of L.M.N. and S.Z. 
was supported by the Q-CAYLE project, funded by the European Union-Next Generation UE/MICIU/Plan de Recuperacion, Transformacion y Resiliencia/Junta de Castilla y Leon (PRTRC17.11), and also by RED2022-134301-T and PID2023-148409NB-I00, financed by MI-CIU/AEI/10.13039/501100011033. Financial support of the Department of Education of the Junta de Castilla y León and FEDER Funds is also gratefully acknowledged (Reference: CLU-2023-1-05).

\end{document}